\documentclass{article}
\usepackage[utf8]{inputenc}
\usepackage{amsmath}
\usepackage{graphicx}

\usepackage{url}
\usepackage{booktabs}
\usepackage{multirow}

\usepackage{amsmath,amsfonts}
\usepackage{algorithmic}
\usepackage{algorithm}
\usepackage{graphicx} 
\usepackage{graphbox}
\usepackage[caption=false,font=normalsize,labelfont=sf,textfont=sf]{subfig}
\usepackage{array}
\usepackage{textcomp}
\usepackage{stfloats}
\usepackage{url}
\usepackage{pgfplots}
\usepackage{multirow}
\usepackage{verbatim}
\usepackage{float}
\usepackage{booktabs}

\title{Batch-oriented Element-wise Approximate Activation for Privacy-Preserving Neural Networks}

\author{
  Peng Zhang\\
  Shenzhen University\\
  \texttt{zhangp@szu.edu.cn}
  \and
  Ao Duan\\
  Shenzhen University\\
  \texttt{2110436077@email.szu.edu.cn}
  \and
  Xianglu Zou\\
  Shenzhen University\\
  \texttt{2060432083@email.szu.edu.cn}
  \and
  Yuhong Liu\\
  Santa Clara University\\
  \texttt{yhliu@scu.edu}
}

\begin{document}
\maketitle
\begin{abstract} 
Privacy-Preserving Neural Networks (PPNN) are advanced to perform inference without breaching user privacy, which can serve as an essential tool for medical diagnosis to simultaneously achieve big data utility and privacy protection. 
As one of the key techniques to enable PPNN, Fully Homomorphic Encryption (FHE) is facing a great challenge that homomorphic operations cannot be easily adapted for non-linear activation calculations.
In this paper, batch-oriented element-wise data packing and approximate activation are proposed, which train linear low-degree polynomials to approximate the non-linear activation function - ReLU. 
Compared with other approximate activation methods, the proposed fine-grained, trainable approximation scheme can effectively reduce the accuracy loss caused by approximation errors. 
Meanwhile, due to element-wise data packing, a large batch of images can be packed and inferred concurrently, leading to a much higher utility ratio of ciphertext slots. Therefore, although the total inference time increases sharply, the amortized time for each image actually decreases, especially when the batch size increases. 
Furthermore, knowledge distillation is adopted in the training process to further enhance the inference accuracy.
Experiment results show that when ciphertext inference is performed on 4096 input images, compared with the current most efficient channel-wise method, the inference accuracy is improved by $1.65\%$, and the amortized inference time is reduced by $99.5\%$.
\end{abstract}



\section{Introduction}
Medical diagnosis is the process of identifying diseases based on users' medical data, which often requires intensive labor work and rich clinical experiences from doctors. It is also possible for doctors to make a misjudgment. With the development of technologies, deep neural networks have been used for medical diagnosis \cite{kaissis2021end} to improve doctors' work efficiency and accuracy, and facilitate early detection and diagnosis of diseases. Since the data analysis process involves a significant amount of computations, neural network based medical diagnosis is often outsourced to external third party service providers. 

However, as the medical data contains user private information, Privacy-Preserving Neural Networks (PPNN) that can achieve high utility while not sacrificing user privacy are required \cite{alex2023energy}. Fully Homomorphic Encryption (FHE)\cite{kim2023optimized} has been considered as one of the key technologies to achieve PPNN, as it supports operations over the encrypted data. In particular, with FHE-enabled PPNN, a user can encrypt his/her private data and only upload the ciphertext to a server for analysis. The server can perform inference over the encrypted data and return encrypted inference results to the user, the only party who can decrypt the inference results. In addition, since the inference model is only hold by the server, both user data privacy and the model security are guaranteed.

However, although most FHE schemes, such as BFV \cite{fan2012somewhat}, BGV \cite{brakerski2014leveled}, CKKS \cite{cheon2017homomorphic}, RNS-CKKS \cite{cheon2019full}, can well support linear calculations (e.g., additions and multiplications) \cite{akavia2022privacy}, they cannot efficiently handle non-linear calculations, such as max-pooling and non-linear activation functions, which are very important components of neural networks. Therefore, the most popular solution is to replace or approximate these non-linear calculations by linear functions. For example, average pooling is often adopted to replace max-pooling, which can generally achieve a good model accuracy. Nevertheless, how to approximate those non-linear activation functions in a linear way without sacrificing much accuracy and efficiency remains a challenge and has attracted extensive research attentions.

One of the most commonly used and stable activation function in neural networks is the ReLU function. 
In the PPNN model first proposed by Gilad et al. \cite{gilad2016cryptonets}, a square activation function $P(x)=x^{2}$ is directly used to replace ReLU activation function. This approximation, however, can only adapt to shallow neural networks with fewer layers. When the neural network layer deepens, the approximate errors will increase rapidly, which may significantly affect the inference accuracy of the neural network.

Typically, there are two ways to improve the inference accuracy. One is using a high-degree polynomial to approximate ReLU. For example, Lee et al. \cite{lee2021precise} uses minimax composite polynomials to approximate ReLU.
This method results in a large multiplication depth, so that time-consuming bootstrapping operations \cite{han2020better} in FHE are required to control noise growth.
While the adoption of a high-degree polynomial can improve the inference accuracy, it also causes a significant increase in inference time.
The second way to improve the accuracy is using trainable polynomial activation, which can adaptively learn the parameters of polynomials during the training process. 
Wu et al. \cite{wu2018ppolynets} proposed a 2-degree polynomial to approximate ReLU. 
After that, SAFENet \cite{lou2021safenet} and HEMET \cite{lou2021hemet} also used trainable polynomial to replace ReLU to implement PPNN.
Polynomial parameters, together with the model parameters, are trained until the loss function converges.
This method can reduce the inference accuracy loss caused by approximation while not incurring high computational cost. Therefore, we follow the second approach in this study. 

Specifically, most existing studies either train one approximate polynomial for all channels, called layer-wise approximate activation \cite{lou2021hemet,dathathri2019chet}; or train one polynomial for each channel to provide further fine-grained approximation, named channel-wise approximate activation \cite{wu2018ppolynets,lou2021safenet}.
Both layer-wise approximate activation and channel-wise approximate activation use the same data packing, where all elements per channel are packed into one. Taking CIFAR-10 as an example, the size of pack is only $32\times32$.
However, such a packing solution does not naturally match the FHE schemes. For example, considering using RNS-CKKS to encrypt the pack, there are $N/2$ ciphertext slots available, where $N$ is the ring polynomial degree. With a typical setting of $N$ value as 32,768, there are 16,384 slots that can be used in theory. It means that the CIFAR-10 data set will only take 6.25\% of the available slots, resulting in a very low utility ratio.

Therefore, in this study, in order to reduce accuracy loss caused by approximation and improve the utility ratio of the ciphertext slots, we propose a Batch-oriented Element-wise Approximate Activation (BEAA) approach for PPNN, which focuses on a more efficient data packing and a finer-grained and trainable approximate activation. Our major contributions are summarized as follows.

\begin{itemize}

\item[$\bullet$] In order to improve the utility ratio of the ciphertext slots, element-wise data packing is proposed, which enables a batch-oriented packing of images with a maximum batch size as $N/2$. It indicates that a large batch of images can be packed together for fully homomorphic encryption and later be inferred concurrently, resulting in a significantly reduced amortized inference time. 

\item[$\bullet$] To improve the model accuracy, an element-wise approximate activation scheme is proposed, which enables fine-grained, trainable approximation functions so that the inference accuracy loss caused by approximation errors is reduced. Furthermore, we propose to train neural networks with the approximate polynomials instead of the original activation functions, and introduce knowledge distillation into the training processing as it can distill the knowledge from a large teacher model into a small student model. 

\item[$\bullet$] A privacy-preserving neural network is implemented, where the network model is an optimized SqueezeNet; the data is encrypted by FHE; and the proposed scheme BEAA is used to approximate ReLU.
CIFAR-10 dataset and a medical dataset OCTID are employed to validate the performance of the proposed scheme. 
Experiment results show that the accuracy of the proposed BEAA is very close to that of the original ReLU.
Knowledge distillation also contributes to improve the accuracy. Although the overall inference time is large, the amortized time is significantly shortened compared to other FHE-enabled PPNN schemes.
\end{itemize}

 \section{Related works}
 When FHE is used for PPNN, as FHE only supports linear computations over the encrypted data, how to approximate non-linear activation functions such as ReLU has been one of the key points. 
 
 Various approximation through low-degree polynomials have been proposed. For example, CryptoNets \cite{gilad2016cryptonets} initially used square polynomial $x^2$ to replace the activation function, followed by LoLa \cite{brutzkus2019low} and Delphi \cite{mishra2020delphi}. 
 Faster CryptoNets \cite{chou2018faster} also tried to approximate activation using the low-degree polynomial $2^{-3}x^2+2^{-1}x+2^{-2}$. 
 Although low-degree polynomial approximation can achieve high efficiency, it often causes large accuracy loss, especially when adopted by deep neural networks.
 
 In order to reduce the accuracy loss, Chabanne et al. \cite{chabanne2017privacy} used Taylor series to approximate ReLU activation function. CryptoDL \cite{hesamifard2017cryptodl} used the Chebyshev polynomials to approximate the derivative of the activation function, so that the approximate polynomial of the activation function can be retrieved. 
 Podschwadt et al. \cite{podschwadt2020classification} approximated the Tanh function with 3-degree polynomials for RNN. 
 Subsequently, Lee et al. \cite{lee2021precise} used the minimax composite polynomial to approximate ReLU to achieve high accuracy.
 These researches show that using high-degree polynomials to approximate the activation functions can reduce accuracy loss. 
 However, high-degree polynomials will produce a large amount of computations and consume the depth of circuits rapidly, which results in additional bootstrapping operations of FHE to control the decryption noises.

 To reduce the accuracy loss while achieving high efficiency, Chabanne et al. \cite{chabanne2017privacy} proposed to train neural networks with the 
 approximate polynomials instead of original activation functions.
 HEMET \cite{lou2021hemet} used the polynomial $ax^2+bx+c$. 
 Similarly, CHET \cite{dathathri2019chet} also used trainable 2-degree polynomials as their activation functions. 
 After that, Jang et al. \cite{jang2022privacy} attempted to use a polynomial of degree 7 with trainable coefficients. However, this scheme requires bootstrapping and is thus less feasible.

The above trainable polynomials are coarse-grained, where the approximation of the polynomial is layer-wise type approximation. Using the same approximate polynomial for all data in the same layer will cause a large loss of accuracy. Wu et al.  \cite{wu2018ppolynets} proposed a channel-wise approximate activation scheme. They utilized a 2-degree polynomial approximation of the form $ax^2+bx$ for the activation function, where $a$ and $b$ are trainable parameters. SAFENet \cite{lou2021safenet} later adopted this finer-grained approximation scheme, along with the use of two coefficients trainable polynomials $a_1x^3+a_2x^2+a_3x+a_4$ and $b_1x^2+b_2x+b_3$ to approximate the activation function.
 
 As discussed above, on the one hand, low-degree polynomial approximation is efficient but causing large accuracy loss. On the other hand, high-degree polynomial approximation can improve accuracy but will cause high computational overhead. To balance the trade-off, trainable polynomials are proposed, which can achieve acceptable accuracy loss with reasonable computational overhead. Most schemes adopting trainable polynomials, however, use the same polynomial for all channels. As there are typically multiple channels and hundreds of features in neural networks, in this work, we aim to propose a finer-gained scheme to approximate activation so that the errors introduced by polynomial approximation can be further reduced.

\section{Preliminaries}

\subsection{Fully homomorphic encryption}
 Among various FHE schemes, CKKS \cite{cheon2017homomorphic} can well control the growth of ciphertext size and noise caused by homomorphic multiplication through ciphertext rescaling operation. In addition, it can efficiently perform high-precision fixed-point operations, thus becoming one of the most suitable FHE schemes for PPNN. On the premise of using the Chinese Remainder Theorem (CRT), RNS-CKKS \cite{cheon2019full} can keep the ciphertext in the form of RNS (Residue Number System) variants, avoid the CRT conversion operation of multi-precision integers, and further improve the computational efficiency. 
 Therefore, in this paper, we adopt RNS-CKKS to encrypt the data, which is described as follows.
 
$ParamsGen$: Key parameters are set, including the security parameter $\lambda$, the ring polynomial degree $N$, the coeffcient modulus $q$ and a discrete Gaussian distribution $\chi$.

$KeyGen$: According to the input parameters $(N, q, \chi)$, the following operations are performed. (1) From the sparse distribution $R$, $R_q$, $\chi$ on $\{0, \pm 1\}^N$, we randomly select polynomials $s$, $a$, $e$, respectively, and then generate the private key $sk=(1,s)$ and the public key $pk=(b,a)\in R_q^2$, where $b \leftarrow -as+e(modq)$ ; (2) Let $Q=q^2$, randomly select polynomials $s^{\prime}$, $a^{\prime}$, $e^{\prime}$ from $s^2$, $R_Q$, $\chi$, respectively, and obtain the evaluation key $evk=(b^{\prime}, a^{\prime})\in R_Q^2$, where $b ^{\prime}\leftarrow -a^{\prime}s+e^{\prime}+qs^{\prime}(modQ)$.

$Enc$: For a given plaintext $m$, the polynomial $v$ and error polynomials $e_0$, $e_1$ are extracted from the random distribution $\chi$, and the ciphertext is obtained as: $\textbf{c}=v\cdot pk+(m+e_0,e_1)(modq)$.

$Dec$: For a given ciphertext $\textbf{c}=(c_0, c_1)$, the decrypted plaintext is $m=c_0+c_1s(modq)$.

The main arithmetic operations supported by RNS-CKKS are homomorphic addition $Add$, homomorphic scalar multiplication $CMult$, homomorphic multiplication $Mult$, homomorphic rotation $Rot$, which are discribed as follows. 

$Add$: Given the ciphertext $\textbf{c}_1,\textbf{c}_2$, the output of the homomorphic addition is $\textbf{c}_{add}=\textbf{c}_1+\textbf{c}_2(modq)$. Generally, homomorphic additions can be directly used to replace the addition operations in PPNN, and the overhead is relatively low.

$CMult$: Given the ciphertext $\textbf{c}$ and plaintext $m$, the output of the homomorphic scalar multiplication is $\textbf{c}_{CMult}=m\textbf{c}(modq)$. As the model parameters in PPNN are plaintext, homomorphic scalar multiplications are often used, and the overhead is low.

$Mult$: For two given ciphertexts: $\textbf{c}=(c_0, c_1)$, $ \textbf{c} ^{\prime}=(c_0 ^{\prime}, c_1^{\prime})$, and the evaluation key $evk$, we define $(d_0, d_1, d_2)=(c_0c_0^{\prime}, \ c_0 c_1 ^{\prime}+c_1 c_0 ^{\prime}, \ c_1 c_1^{\prime}) (modq)$. Finally, the result of homomorphic multiplication is $\textbf{c}_{mult}=(d_0, d_1)+\lfloor 1/q \cdot d_2 \cdot evk\rfloor(modq)$. In general, this operation, which is often performed at the activation layer, has a very high overhead.

$Rot$: Given the ciphertext $\textbf{c}$ and the rotation times $k$, this function outputs $\textbf{c}^{\prime}$, which is the ciphertext of the vector $(m_{k+1}, m _{k+2},..., m_l, m_1,..., m_k)$ obtained by shifting the plaintext vector $(m_1, m_2,..., m_l)$ of $\textbf{c}$ to the left by $k$ slots. With a very high cost, the $Rot$ operation is often used with $CMult$ to achieve homomorphic convolution in the convolution layer.
 
\subsection{Layer-wise approximate activation}
\begin{figure}[!t]
   \centering  
    \subfloat[ReLU activation]{\includegraphics[width=0.405\textwidth]{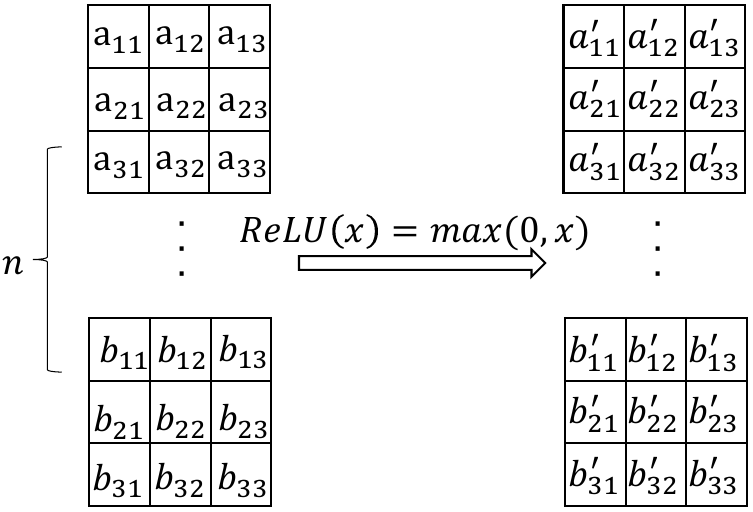}%
    \label{fig:relu}}
     \hfil
    \subfloat[Layer-wise approximate activation]{\includegraphics[width=0.405\textwidth]{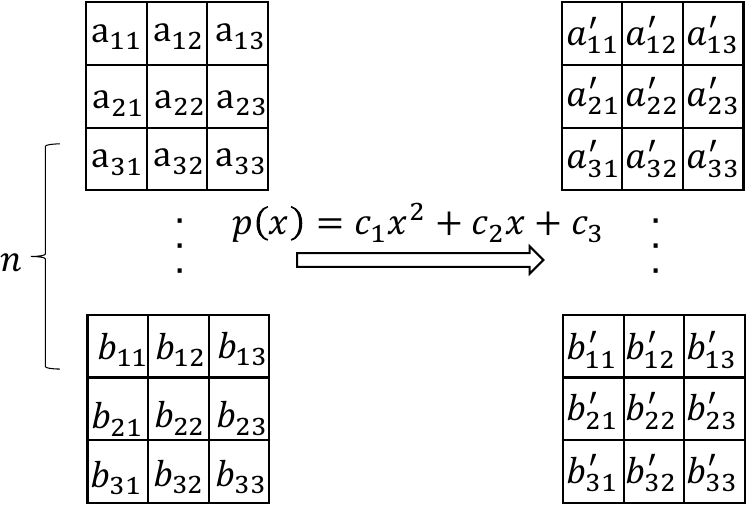}%
    \label{fig:layer-wise1}}
    \caption{ReLU activation and Layer-wise approximate activation.}
    \label{fig:layer}
\end{figure}

As we discussed before, classic activation functions such as ReLU and Sigmoid are non-linear, which cannot be used in PPNN prediction directly. Considering the efficiency, low-degree polynomials are used to approximate the activation functions. If one approximate activation function is used for a convolutional layer, it is called layer-wise approximate activation. An example is shown in Figure \ref{fig:layer}.

 Formally, a 2-degree parametric polynomial is defined as:
   \begin{eqnarray}
    	\begin{array}{rcl}
    		p(x)=c_{1}x^{2}+c_{2}x+c_3
    	\end{array}
    \end{eqnarray}

 where $x$ is a $n\times H\times W$ three-dimensional matrix, representing the input eigenvalues of the current activation layer, generally from the convolution layer or the batch normalization layer; $n$ represents the number of input channels of the current convolution layer; $H$ and $W$ represent the height and width of the image, respectively. $c_1$, $c_2$, $c_3$ represent the parameters to be trained for the quadratic term, the linear term and the constant term of the activation layer, respectively, which can be obtained through iterative training with the back propagation. Please refer to \cite{lou2021hemet} for more details.

\subsection{Channel-wise approximate activation}
\begin{figure}[!t]
   \centering  
    \subfloat[Layer-wise approximate activation]{\includegraphics[width=0.405\textwidth]{layer-wise.pdf}%
    \label{fig:layer-wise2}}
     \hfil
    \subfloat[Channel-wise approximate activation]{\includegraphics[width=0.52\textwidth]{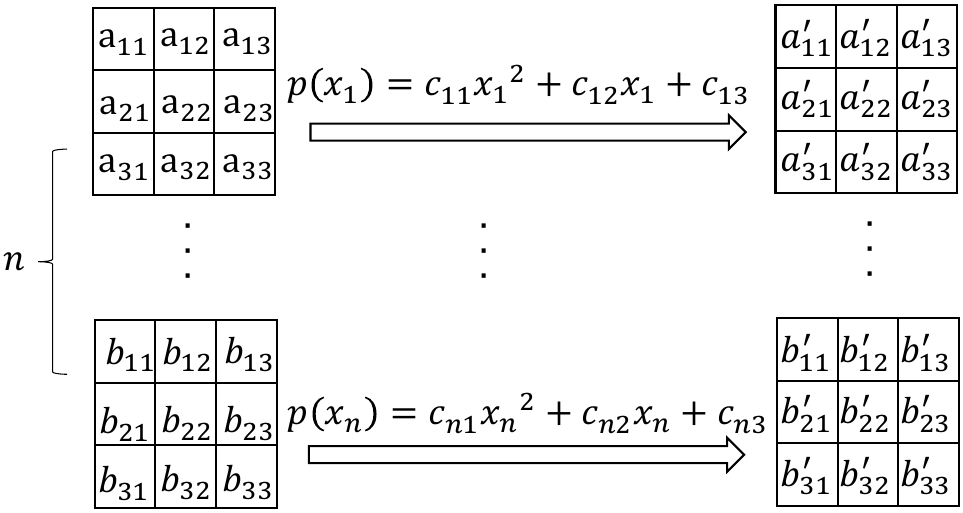}%
    \label{fig:channel-wise}}
    \caption{Layer-wise and Channel-wise polynomial approximation.}
    \label{fig:channel}
\end{figure}

Using one polynomial activation in a layer will inevitably result in some errors, even if the polynomial is parametric. As there are $n$ channels in the convolution layerm, if the approximate polynomials on different channels are different, the approximate error may be controlled. Therefore, channel-wised approximate activation is proposed by Wu et al. \cite{wu2018ppolynets}. As shown in Figure \ref{fig:channel}, for the input eigenvalue $x_i$ on the $i$-th channel, the activation polynomial is $p(x_i) (i=1,\cdots,n)$. 

 \noindent 
 For any activation layer with $n$ input channels, the activation function is defined as:
\begin{eqnarray}
    \begin{array}{rcl}
        p(x_i)=c_{i1}x_{i}^{2}+c_{i2}x_{i}+c_{i3} (i=1,\cdots,n)
    \end{array}
\end{eqnarray}
where $x_i$ represents the input eigenvalue of $i$-th channel, which is usually a  $H\times W$ two-dimensional matrix. In addition, $c_{i1}$, $c_{i2}$ and $c_{i3}$ represent the parameters to be trained for the quadratic, linear and constant terms of the polynomial for the $i$-th input channel, respectively, which can be obtained by iterative training through the back propagation. Please refer to \cite{wu2018ppolynets} for more details.
 
\section{Batch-oriented element-wise approximate activation}
As discussed above, using low-degree polynomials to approximate ReLU activation function will inevitably incur approximate errors; and channel-wise approximate activation is expected to reduce the approximate error. 
To further reduce the approximate error, based on channel-wise approximate activation, a finer-grained scheme, namely Batch-oriented Element-wise Approximate Activation (BEAA), is proposed in this section, where each feature element in the activation layer is separately trained. 

\subsection{Data packing}
In the layer-wise and channel-wise polynomial approximation, the typical data packing method is used, namely, channel-wise data packing. 
Given $M$ plaintext input images, each image is converted into a $n\times H\times W$ matrix, where $n$ represents the number of input channels, and $H$ and $W$ represent the height and width of the image, respectively.
Next, all images are packed into $M\times n$ plaintext, and each plaintext contains $H\times W$ feature elements. 
And then, the plaintext is encrypted to obtain $M\times n$ ciphertext by RNS-CKKS. By packing all the $H\times W$ elements within one channel together, this method cannot support further processing at a finer-grained level (e.g. element-wise approximation).

\begin{figure}[!t]
\centering  
\noindent
\includegraphics[ width=0.5\textwidth]{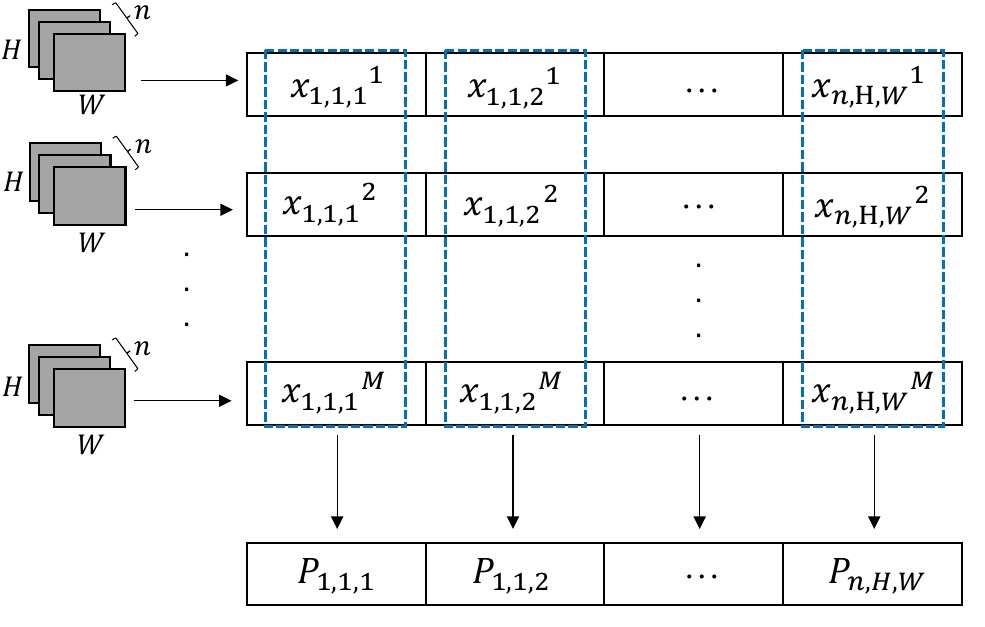}
\caption{Element-wise data packing.}
\label{fig:packaging}
\end{figure}

Therefore, in this work, we propose to perform an element-wise data packing, which has been proved to be an effective way to promote homomorphic SIMD (Single Instruction Multiple Data) in privacy-preserving neural network schemes, such as CryptoNets \cite{gilad2016cryptonets} and  nGraph-HE \cite{boemer2019ngraph}. 
As shown in Figure \ref{fig:packaging}, given $M$ plaintext input images, each image consists of $n\times H\times W$ features. Element-wise data packing scheme packs the features at the corresponding position of each image into a plaintext $P_{n,H,W}$, where each plaintext contains $M$ feature data integrated from the $M$ input images.
The plaintext $P_{n,H,W}$ is then encrypted by RNS-CKKS to obtain the corresponding ciphertext $X_{n,H,W}$.

Furthermore, since RNS-CKKS can encrypt multiple messages into a single ciphertext, the number of packed messages can be adjusted based on the number of ciphertext slots. For example, when the ring polynomial degree $N$ is set as 32,768, there are 16,384 slots available. 
That means, a large batch of images can be dealt concurrently, without introducing additional time cost.

\subsection{Proposed approximate algorithm}
\begin{figure}[!t]
\centering  
\includegraphics[width=0.5\textwidth]{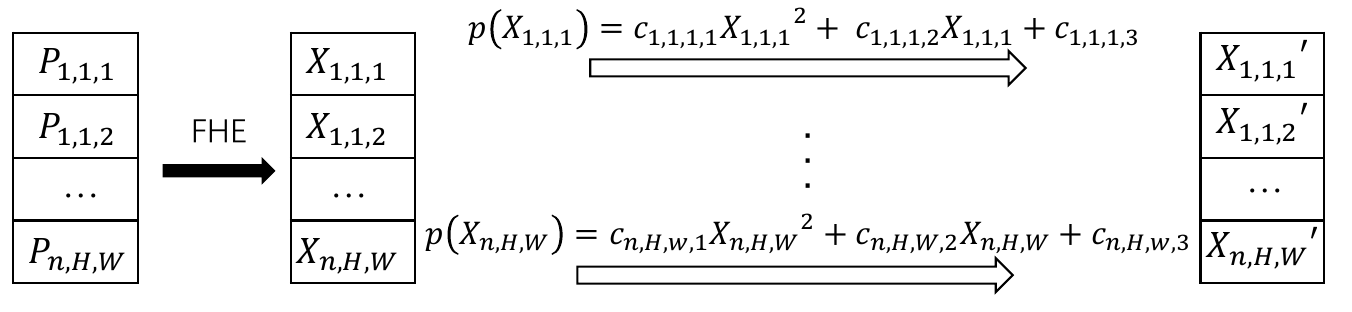}
 \caption{Element-wise approximate activation.}
 \label{fig:pro_scheme}
\end{figure}

In the proposed approximate scheme BEAA, each ciphertext is approximately activated by using 2-degree polynomials instead of ReLU. More specifically, the approximation is performed at the element level, indicating that each single feature element is approximated by one polynomial. As shown in Figure \ref{fig:pro_scheme}, for the feature element $X_{i,h,w}$ in the $h^{th}$ row and $w^{th}$ column of the feature matrix of the $i^{th}$ channel, the activation polynomial is $p(X_{i,h,w})$.

For a given feature element $X_{i,h,w}$, the activation function is defined as:
\begin{eqnarray}
    	\begin{array}{rcl}
    		p(X_{i,h,w})=c_{i,h,w,1}X_{i,h,w}^2+c_{i,h,w,2}X_{i,h,w}+c_{i,h,w,3}
    	\end{array}
    \end{eqnarray}
where $c_{i,h,w,1}$, $c_{i,h,w,2}$, $c_{i,h,w,3}$ represent the parameters to be trained for the quadratic term, the linear term and the constant term of the approximate activation function, respectively. The training is performed through back propagation algorithm. Firstly, the partial derivative is obtained:
 \begin{eqnarray}
    	\begin{array}{rcl}
    		\frac{\partial p(X_{i,h,w})}{\partial c_{i,h,w,1}}=X_{i,h,w}^2, \frac{\partial p(X_{i,h,w})}{\partial c_{i,h,w,2}}=X_{i,h,w}, \frac{\partial p(X_{i,h,w})}{\partial c_{i,h,w,3}}=0
    	\end{array}
    \end{eqnarray}

    \noindent According to the chain rule, the gradient of $c_{i,h,w,j} (j\in\{1,2,3\})$ can be obtained:
    \begin{eqnarray}
    	\begin{array}{rcl}
    		\nabla J(c_{i,h,w,j})=\sum_{X_{i,h,w}} \frac{\partial J(c_{i,h,w,j})}{\partial p(X_{i,h,w})} \frac{\partial p(X_{i,h,w})}{\partial c_{i,h,w,j}}
    	\end{array}
    \end{eqnarray}

    \noindent where $J(c_{i,h,w,j})$ represents the objective loss function of the neural network, and $\frac{\partial J(c_{i,h,w,j})}{\partial p(X_{i,h,w})}$ represents the gradient from the back propagation of the deeper network. Then $L_2$ regularization \cite{cortes2012l2} is also introduced to reduce the over-fitting in the training process, so the objective function $J(c_{i,h,w,j})$ is adjusted to :
\begin{eqnarray}
    	\begin{array}{rcl}
    	 J(c_{i,h,w,j}) =J(c_{i,h,w,j})+\frac{\lambda}{2}J(c_{i,h,w,j})^{2}=\sum_{X_{i,h,w}} \frac{\partial J(c_{i,h,w,j})}{\partial p(X_{i,h,w})} \frac{\partial p(X_{i,h,w})}{\partial c_{i,h,w,j}}+\frac{\lambda}{2}J(c_{i,h,w,j})^{2}
    	\end{array}
    \end{eqnarray}
where $\lambda$ is the attenuation coefficient of the regular term. Therefore, the gradient is updated to:
\begin{eqnarray}
    	\begin{array}{rcl}
    		\nabla J(c_{i,h,w,j})
      =\sum_{X_{i,h,w}} \frac{\partial J(c_{i,h,w,j})}{\partial p(X_{i,h,w})} \frac{\partial p(X_{i,h,w})}{\partial c_{i,h,w,j}}
      +\lambda c_{i,h,w,j}
    	\end{array}
    \end{eqnarray}
Finally, the Nesterov gradient descent algorithm \cite{ruder2016overview} is used to iteratively update the parameter  $c_{i,h,w,j}(j\in\{1,2,3\})$ of the activation function in the network :
\begin{eqnarray}
    	\begin{array}{rcl}
    		V^{t}_{c_{i,h,w,j}}=\mu V^{t-1}_{c_{i,h,w,j}}-\alpha \nabla J(c_{i,h,w,j}^t+\mu V^{t-1}_{c_{i,h,w,j}})
    	\end{array}
    \end{eqnarray}
    \begin{eqnarray}
    	\begin{array}{rcl}
    		c_{i,h,w,j}^t=c^{t-1}_{i,h,w,j}+V^{t-1}_{c_{i,h,w,j}}
    	\end{array}
    \end{eqnarray}
where $V^{t}_{c_{i,h,w,j}}$ represents the impulse during the gradient descent training of $c_{i,h,w,j}$ parameter at time $t$, which is initialized to 0. In addition, $\mu$, $\alpha$ are two hyperparameters, representing the impulse coefficient and learning rate, respectively.

For a single image, which consists of $n\times H\times W$ features, the number of parameters introduced by BEAA is $3nHW$. Compared to 3 parameters for layer-wise approximate activation, and $3n$ parameters for the element-wise approximate activation, the number of parameters for the element-wise approximate activation increases sharply, leading to significant increase in training and inference time. Thanks to the element-wise data packing, which concurrently packs large batch images, BEAA can perform inference on large batch encrypted images in parallel, so that the share-out cost for one single image is competitive. 
    
\section{Activation polynomial training based on knowledge distillation}

The activation polynomials in BEAA are parametric, which are obtained from data training. To further improve the approximate accuracy, knowledge distillation technology is introduced for the training processing. 
Specifically, the mechanism of knowledge distillation \cite{wang2018adversarial,baruch2022methodology} is that the student model learns the knowledge output of the teacher model through knowledge distillation to improve its generalization ability and prediction accuracy.

In this paper, the teacher model refers to the optimized SqueezeNet in HEMET \cite{lou2021hemet} with two Fire modules using ReLU activation function. 
The teacher model is the normal training of our model in plaintext. 
After the data is calculated by each layer of neural networks, the probability of the training data for each classification label is predicted by the $Softmax$ function. 
By comparing the classification results with the original labels, the final loss of the neural network prediction can be obtained.
And then, the model parameters are iteratively trained according to back propagation and gradient descent algorithm.

\begin{figure}[!t]
\centering  
\includegraphics[ width=\textwidth]{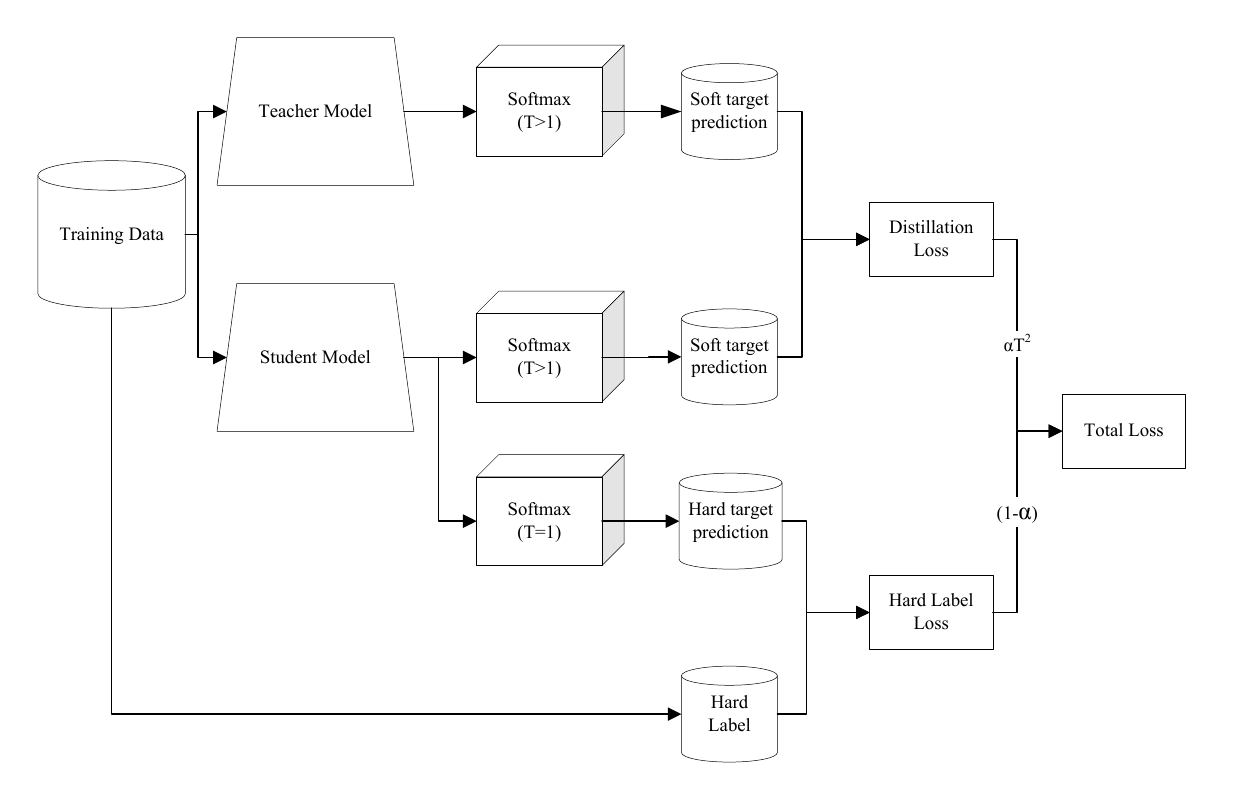}
 \caption{The model training with knowledge distillation.}
 \label{fig:kd}
\end{figure}

The obtained teacher model after training is used to guide the training of the student model. The knowledge of the teacher model is transmitted to the student model in the form of distillation loss. 
In this paper, the student model refers to the optimized SqueezeNet that uses trainable polynomials instead of ReLU.
As shown in Figure \ref{fig:kd}, both the soft target and the hard label (real label) are used to instruct the training of the student model, so that the generalization performance and prediction accuracy of the student model are improved.

Knowledge distillation controls the uniformity of the output probability distribution by introducing a temperature coefficient $T$ in the $Softmax$ layer. When $T=1$, the hard target prediction represents the probability distribution output using the standard $Softmax$ function, and the cross entropy between the hard target prediction and the hard label is called the hard label loss. 
When  $T \textgreater 1$, the probability distribution of the $Softmax$ function output will be more uniform, and this output is called soft target prediction. The cross entropy between the soft target predictions of the teacher model and that of the student model is called the distillation loss. 
The final loss is a weighted average of the two losses, which is computed as the sum of $\alpha T^2$ times the distillation loss and $(1-\alpha)$ times the hard label loss, where the parameter $\alpha$ represents the distillation intensity.

\section{Experiments and analysis}

\subsection{Experiment setup}
{\bf{Environments}}. In this work, one server is involved for all experimental tests. Specifically, the CPU of the server is Intel (R) Core (TM) i9-10900X with 10 cores and 256GB RAM memory.
Neural networks are trained using the TensorFlow framework \cite{abadi2016tensorflow}. Based on the Microsoft SEAL library \cite{sealcrypto}, the data is encrypted by the FHE algorithm RNS-CKKS \cite{cheon2019full}. In order to ensure that all ciphertext can reach the 128-bit security level in PPNN, the ring polynomial degree $N$ is 32,768, and the coeffcient modulus $q$ is 720-bit.

\noindent {\bf{Dataset}}. The CIFAR-10 dataset \cite{torralba200880} and OCTID dataset\cite{gholami2020octid} were used during the experiments. 
CIFAR-10 is a data set with ten categories, a total of 60,000 images, including 10,000 test data sets, and the size of each image is $3\times 32\times 32$.
The OCTID dataset is a cross-sectional image of the retina obtained by Optical Coherence Tomography (OCT). A total of 572 images were included, including 208 normal (NORMAL) retinal images and 4 different retinopathy scanning image samples: 57 Age-related Macular Degeneration (AMD) images, 104 Central Serous Retinopathy (CSR) images, 109 Diabetic Retinopathy (DR) images, and 104 Macular Hole (MH) images. Before the experiments, all images were adjusted from the original $500\times 750$ features to $112\times 112$ features. Due to the small sample size in the data set, in order to avoid overfitting linearity, we processed the images in the data set by rotation, cutting, scaling and other processing to increase the number of samples and improve the generalization performance of the model. Finally, 22800 processed images were obtained. 60$\%$ of the final data set is used for training, 10$\%$ for validity testing, and 30$\%$ for testing. Different types of pathological images are evenly distributed to the training, verification and test data sets to ensure good data distribution.  

\noindent {\bf{Data Packing}}. 
Given $M$ plaintext input images, assume that each image is composed of $H\times W$ features; and the number of channels for each image is $n$.
The element-wise data packing performed for element-wise approximate activation, which packs the features of a specific channel in a large batch of images together to form a plaintext.
Thus, $H\times W\times C$ data can be obtained.
In addition, the channel-wise data packing method, which packs each channel of each image into a ciphertext, is used for both layer-wise and channel-wise approximate activation. 
Thus, $M\times n$ ciphertext data can be obtained. 
Then, the fully homomorphic encryption algorithm RNS-CKKS is used to encode and encrypt the packed data to obtain the ciphertext data. 

\begin{figure}[!t]
    \centering  
       
    \subfloat[Network Model]{\includegraphics[ width=1\textwidth]{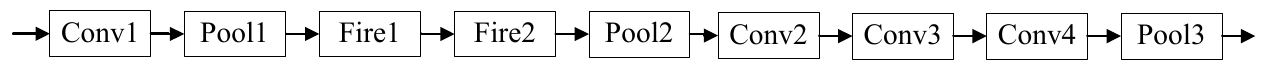}%
    \label{fig:network_model}}\\
    
    \subfloat[Conv Model]{\includegraphics[width=0.65\textwidth]{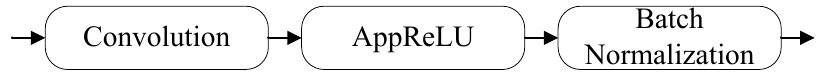}%
    \label{fig:conv_model}}
    
    \caption{Optimized SqueezeNet model and its internal modules.}
    \label{fig:network}

\end{figure}

\noindent {\bf{Neural Networks}}. The network model adopted in this work is optimized SqueezeNet in HEMET \cite{lou2021hemet}, as shown in Figure \ref{fig:network_model}. It is an optimized model structure from SqueezeNet, with four layers of Conv module, two layers of Fire module and three layers of Pool module. The structure of each Conv module is shown in Figure \ref{fig:conv_model}, including a convolution layer, an approximate activation layer and a batch normalization layer. 
The number of output channels of the Conv4 module determines the final output channel number of the model. The setting of the output channel number is related to the experimental task and the data type. If it is a 10-category task, the number of output channels is set to 10, and if it is a 5-category task, the number of output channels is adjusted to 5.
The Fire module consists of two distinct types of convolutional layers: the squeeze layer and the expand layer. The squeeze layer utilizes smaller $1\times1$  convolutional kernels to compress the channel dimension of the input feature maps, thereby reducing computational costs. Following the squeeze layer is the expand layer, which is composed of $1\times1$ and $3\times3$  convolutional layers. The $1\times1$ convolutional layer is responsible for expanding the channel dimension of the input feature maps, while the $3\times3$ convolutional layer performs spatial convolutions along the channel dimension to enhance feature representation.
The Pool module in the network uses the average pooling. In particular, Pool3 uses global average pooling, which computes the average value of the entire input feature map.  

\noindent {\bf{Homomorphic Privacy-Preserving}}. As discussed before, the packed data is encrypted by fully homomorphic encryption algorithm RNS-CKKS. The network training is executed over the plaintext, and the inference is executed over the encrypted data. Therefore, the above Conv module, Fire module and Pool module are performed homomorphically. Except the activation layers, the other layers refer to linear computation, which can be realized by homomorphic addition $Add$, homomorphic scalar multiplication $CMult$, homomorphic multiplication $Mult$ and homomorphic rotation $Rot$ discussed in Section \uppercase\expandafter{\romannumeral 3}.A. As the activation layers are non-linear, homomorphic operations cannot be used directly. Therefore, the activation polynomials after training in Section \uppercase\expandafter{\romannumeral 5} are used to approximate ReLU.

\subsection{Analysis}
The inference time and accuracy of optimized SqueezeNet in Figure \ref{fig:network} are compared and analyzed, when different approximate activation methods are used. 
As existing schemes use different neural networks for different classification tasks, we adopted the layer-wise method in \cite{lou2021hemet,dathathri2019chet} and channel-wise method in \cite{wu2018ppolynets,lou2021safenet} to approximate ReLU in optimized SqueezeNet for comparison.
The activation functions used in the model include: ReLU, 2-degree trainable polynomials by layer-wise, channel-wise and element-wise approximations. 
Two data training methods are: with and without Knowledge Distillation (KD).
All experiments are repeated 10 times. After removing the largest and the smallest, the remaining experimental results are averaged.

Firstly, the inference experiments of CIFAR-10 data set are carried out. 
When a large batch of images are packed by the element-wise data packing method, the batch size is 4096.
As shown in Table \ref{tab:cifar10}, the accuracy of the model activated by ReLU reaches about 87.3$\%$. When approximate activation is used, the inference accuracy decreases  on different degrees.
When layer-wise, channel-wise and element-wise approximate activation are used, the inference accuracy of the PPNN gradually increases, indicating that the finer the granularity is, the less approximate error is.
Based on the channel-wise and element-wise schemes, knowledge distillation assisting activation function training can further improve the inference accuracy. 
In total, with element-wise approximate activation and knowledge distillation, the accuracy enhances 
by $1.65\%$ compared with the current most efficient channel-wise method.

 \begin{table}[!t]
	\centering
	\caption{The inference accuracy and time of optimized SqueezeNet using different approximate activation methods over the encrypted CIFAR-10 dataset.}
	\label{tab:cifar10}  
	\begin{tabular}{ccc}	
        \hline
		Scheme & Inference time (s) & Inference accuracy  \\
		\hline
		ReLU  & $0.43$ & $ 87.28\% $ \\
        Layer-wise \cite{lou2021hemet,dathathri2019chet} & $159.8$ & $ 83.35\%  $ \\
		Channel-wise \cite{wu2018ppolynets,lou2021safenet}& $160.2$ & $ 84.76\%  $ \\
        \multirow{2}*{Element-wise (BEAA)} & $3131.5$(Total) & \multirow{2}*{$ 85.17\%  $} \\
         &$0.764$ (Amortized)&  \\
        
        Channel-wise with KD  & $159.6$ & $ 86.23\% $ \\
        
         \multirow{2}*{Element-wise with KD} & $3134.7$ (Total) & \multirow{2}*{$ 86.41\%$} \\
         &$0.765$ (Amortized)&  \\
    
		\hline
	\end{tabular}
\end{table}

In terms of inference time, it is much longer in PPNN than in plaintext ReLU, as inferences are carried out over the ciphertext encrypted by FHE. 
When using layer-wise and channel-wise approximate activation, the inference time per one image is about 160 seconds. After using BEAA, the inference time is significantly longer, exceeding 3130 seconds. 
Fortunately, it is for a large batch of images (i.e., 4096 images) in the experiments. The inference time for each image is relatively low, only about $0.5\%$ compared with the current most efficient channel-wise method.

Then the inference experiments of OCTID dataset are carried out. Similarly, the selected batch size is 4096. Experimental results are shown in Table \ref{tab:OCTID}, which are consistent with Table \ref{tab:cifar10}.  Compared with CIFAR-10 dataset, the inference accuracy of each class is higer, as OTCID is a 5-category dataset.

\begin{table}[!t]
	\centering
	\caption{The inference accuracy and time of optimized SqueezeNet using different approximate activation methods over the encrypted OCTID dataset.}
	\label{tab:OCTID}  
	\begin{tabular}{ccc}
		
        \hline
		Scheme & Inference time(s)& Inference accuracy  \\
		\hline
		ReLU  & $0.85$  & $ 93.51\%  $ \\
        Layer-wise \cite{lou2021hemet,dathathri2019chet} & $306.3$ & $ 89.42\% $ \\
		Channel-wise \cite{wu2018ppolynets,lou2021safenet}& $306.6$  & $ 90.83\% $ \\
  \multirow{2}*{Element-wise (BEAA)} & $5743.8$ (Total) & \multirow{2}*{$ 91.24\%  $} \\
         &$1.40$ (Amortized)&  \\
        Channel-wise with KD  & $306.5$  & $ 92.36\% $ \\
        \multirow{2}*{Element-wise with KD} & $5741.3$ (Total) & \multirow{2}*{$ 92.47\%$} \\
         &$1.40$ (Amortized)&  \\
		\hline
	\end{tabular}
\end{table}

\begin{figure}[!t]
    \centering  
    \subfloat[The total inference time]{\includegraphics[width=0.5\textwidth]{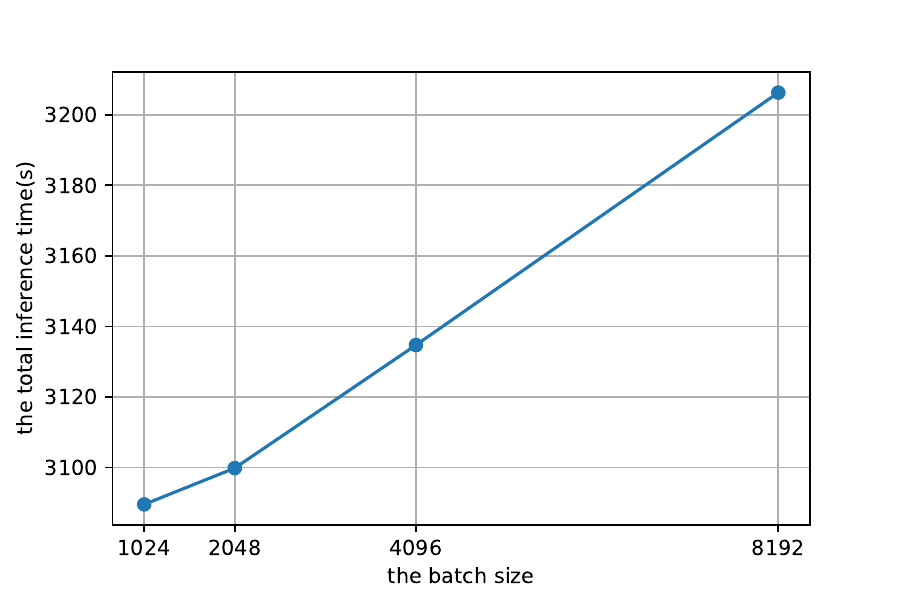}%
    \label{fig:total_time}}
     \hfil
    \subfloat[The amortized inference time]{\includegraphics[width=0.5\textwidth]{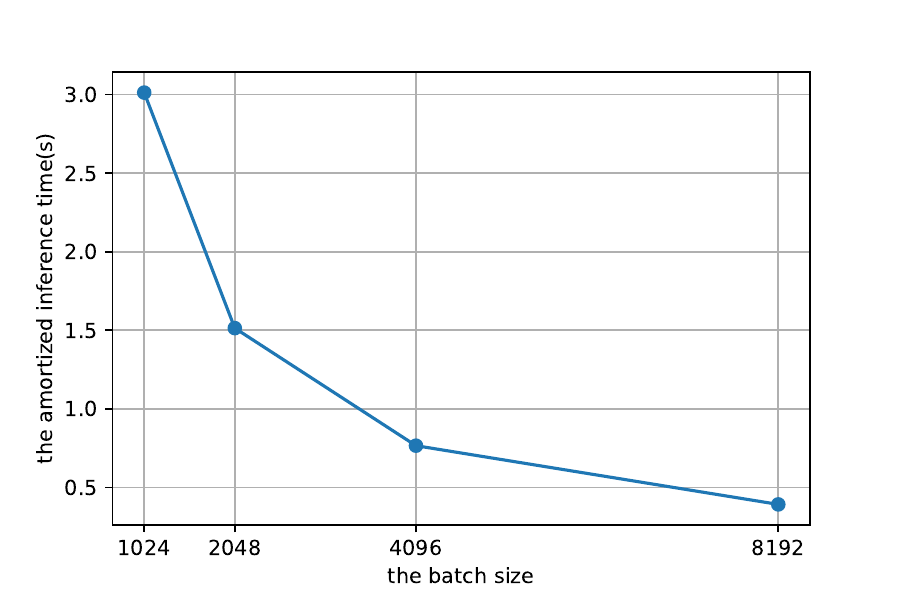}%
    \label{fig:amoritized_time}}
    \caption{The total and amortized inference time when the batch size increases.}
    \label{fig:batch} 
\end{figure}

In theory, as RNS-CKKS supports N/2 ciphertext slots, when the size of pack is no more than N/2, the total inference time for element-wise approximate activation is roughly the same. We test the total inference time when the batch size is 1024, 2048, 4096 and 8192, as shown in Figure \ref{fig:total_time}. There is a small range of time difference, mainly due to the time on reading in the large batch of images.
However, the amortized time for each image decreases with the increase of the batch size, as shown in Figure \ref{fig:amoritized_time}. Obviously, when the batch size is bigger, the utility ratio of ciphertext slots is higher. The best amortized time can be obtained when the slots are fully occupied. Compared with other approximate activation methods, the inference time for each image is proved to be highly competitive.

Given a single image, assume that it consists of $n\times H\times W$ features, then the number of parameters introduced by element-wise approximate activation is $3nHW$, which is high when compared to $3$ parameters in layer-wise approximate activation, and $3n$ parameters in channel-wise approximate activation. 
The number of activation functions of the element-wise approximate activation is the same as the number of output features (i.e., $nHW$), which is also high when compared with that in layer-wise and channel-wise approximate activation (i.e., $n$). 
However, the polynomial parameters are usually trained offline, thus these shortages are tolerable, especially considering that a large batch of images are inferred at the same time, leading to a much shorter amortized inference time for each image (i.e., 0.765s for the CIFAR-10 dataset and 1.4s for the OCTID dataset).

\begin{table}[!t]
	\centering
	\caption{The number of parameters and activation calculations using different approximate activation methods.}
	\label{tab:1}  
	\begin{tabular}{ccc}	
        \hline
		Scheme & Number of parameters & Number of activation calculations  \\
		\hline
		Layer-wise  & $3$ & $n$ \\
		Channel-wise & $3n$ & $n$  \\
            Element-wise & $3n HW$ & $n HW$  \\
		\hline
	\end{tabular}
\end{table}

\section{Conclusion}
With deep neural networks widely adopted by numerous applications for medical data analysis, significant concerns arises around user privacy. As a result, Privacy-Preserving Neural Networks (PPNN) have attracted extensive research attention. By allowing computations to be performed directly on encrypted data, Fully Homomorphic Encryption (FHE) has been recognized as a key technique to enable PPNN. However, as FHE does not support non-linear computations, most existing FHE-enabled PPNN studies adopt polynomials to approximate non-linear functions, which often leads to inference accuracy drops due to approximation errors. In order to improve the model accuracy, a Batch-oriented Element-wise Approximate Activation (BEAA) scheme is proposed in this paper, where each feature element in the activation layer is separately trained by one polynomial. When compared to state-of-the-art FHE-enabled PPNN studies, this finer-grained approximation improves the inference accuracy by $1.65\%$. In addition, as the proposed scheme supports concurrent inference for a large batch of images, the amortized inference time for each image is very competitive, which takes only $0.5\%$ of the inference time of the current most efficient channel-wise approach. Furthermore, when the batch size increases, the utility ratio of ciphertext slots is improved, resulting in a further reduced amortized inference time.

\bibliographystyle{plain}
\bibliography{ref}

\begin{thebibliography}{10}

\bibitem{abadi2016tensorflow}
Mart{\'\i}n Abadi, Ashish Agarwal, Paul Barham, Eugene Brevdo, Zhifeng Chen, Craig Citro, Greg~S Corrado, Andy Davis, Jeffrey Dean, Matthieu Devin, et~al.
\newblock Tensorflow: Large-scale machine learning on heterogeneous distributed systems.
\newblock {\em arXiv preprint arXiv:1603.04467}, 2016.

\bibitem{akavia2022privacy}
Adi Akavia, Max Leibovich, Yehezkel~S Resheff, Roey Ron, Moni Shahar, and Margarita Vald.
\newblock Privacy-preserving decision trees training and prediction.
\newblock {\em ACM Transactions on Privacy and Security}, 25(3):1--30, 2022.

\bibitem{alex2023energy}
Sona Alex, Dhanaraj KJ, and Deepthi PP.
\newblock Energy efficient and secure neural network--based disease detection framework for mobile healthcare network.
\newblock {\em ACM Transactions on Privacy and Security}, 26(3):1--27, 2023.

\bibitem{baruch2022methodology}
Moran Baruch, Nir Drucker, Lev Greenberg, and Guy Moshkowich.
\newblock A methodology for training homomorphic encryption friendly neural networks.
\newblock In {\em Applied Cryptography and Network Security Workshops: ACNS 2022 Satellite Workshops, AIBlock, AIHWS, AIoTS, CIMSS, Cloud S\&P, SCI, SecMT, SiMLA, Rome, Italy, June 20--23, 2022, Proceedings}, pages 536--553. Springer, 2022.

\bibitem{boemer2019ngraph}
Fabian Boemer, Yixing Lao, Rosario Cammarota, and Casimir Wierzynski.
\newblock ngraph-he: a graph compiler for deep learning on homomorphically encrypted data.
\newblock In {\em Proceedings of the 16th ACM International Conference on Computing Frontiers}, pages 3--13, 2019.

\bibitem{brakerski2014leveled}
Zvika Brakerski, Craig Gentry, and Vinod Vaikuntanathan.
\newblock (leveled) fully homomorphic encryption without bootstrapping.
\newblock {\em ACM Transactions on Computation Theory (TOCT)}, 6(3):1--36, 2014.

\bibitem{brutzkus2019low}
Alon Brutzkus, Ran Gilad-Bachrach, and Oren Elisha.
\newblock Low latency privacy preserving inference.
\newblock In {\em International Conference on Machine Learning}, pages 812--821. PMLR, 2019.

\bibitem{chabanne2017privacy}
Herv{\'e} Chabanne, Amaury De~Wargny, Jonathan Milgram, Constance Morel, and Emmanuel Prouff.
\newblock Privacy-preserving classification on deep neural network.
\newblock {\em Cryptology ePrint Archive}, 2017.

\bibitem{cheon2019full}
Jung~Hee Cheon, Kyoohyung Han, Andrey Kim, Miran Kim, and Yongsoo Song.
\newblock A full rns variant of approximate homomorphic encryption.
\newblock In {\em Selected Areas in Cryptography--SAC 2018: 25th International Conference, Calgary, AB, Canada, August 15--17, 2018, Revised Selected Papers 25}, pages 347--368. Springer, 2019.

\bibitem{cheon2017homomorphic}
Jung~Hee Cheon, Andrey Kim, Miran Kim, and Yongsoo Song.
\newblock Homomorphic encryption for arithmetic of approximate numbers.
\newblock In {\em Advances in Cryptology--ASIACRYPT 2017: 23rd International Conference on the Theory and Applications of Cryptology and Information Security, Hong Kong, China, December 3-7, 2017, Proceedings, Part I 23}, pages 409--437. Springer, 2017.

\bibitem{chou2018faster}
Edward Chou, Josh Beal, Daniel Levy, Serena Yeung, Albert Haque, and Li~Fei-Fei.
\newblock Faster cryptonets: Leveraging sparsity for real-world encrypted inference.
\newblock {\em arXiv preprint arXiv:1811.09953}, 2018.

\bibitem{cortes2012l2}
Corinna Cortes, Mehryar Mohri, and Afshin Rostamizadeh.
\newblock L2 regularization for learning kernels.
\newblock {\em arXiv preprint arXiv:1205.2653}, 2012.

\bibitem{dathathri2019chet}
Roshan Dathathri, Olli Saarikivi, Hao Chen, Kim Laine, Kristin Lauter, Saeed Maleki, Madanlal Musuvathi, and Todd Mytkowicz.
\newblock Chet: an optimizing compiler for fully-homomorphic neural-network inferencing.
\newblock In {\em Proceedings of the 40th ACM SIGPLAN Conference on Programming Language Design and Implementation}, pages 142--156, 2019.

\bibitem{fan2012somewhat}
Junfeng Fan and Frederik Vercauteren.
\newblock Somewhat practical fully homomorphic encryption.
\newblock {\em Cryptology ePrint Archive}, 2012.

\bibitem{gholami2020octid}
Peyman Gholami, Priyanka Roy, Mohana~Kuppuswamy Parthasarathy, and Vasudevan Lakshminarayanan.
\newblock Octid: Optical coherence tomography image database.
\newblock {\em Computers \& Electrical Engineering}, 81:106532, 2020.

\bibitem{gilad2016cryptonets}
Ran Gilad-Bachrach, Nathan Dowlin, Kim Laine, Kristin Lauter, Michael Naehrig, and John Wernsing.
\newblock Cryptonets: Applying neural networks to encrypted data with high throughput and accuracy.
\newblock In {\em International conference on machine learning}, pages 201--210. PMLR, 2016.

\bibitem{han2020better}
Kyoohyung Han and Dohyeong Ki.
\newblock Better bootstrapping for approximate homomorphic encryption.
\newblock In {\em Topics in Cryptology--CT-RSA 2020: The Cryptographers’ Track at the RSA Conference 2020, San Francisco, CA, USA, February 24--28, 2020, Proceedings}, pages 364--390. Springer, 2020.

\bibitem{hesamifard2017cryptodl}
Ehsan Hesamifard, Hassan Takabi, and Mehdi Ghasemi.
\newblock Cryptodl: Deep neural networks over encrypted data.
\newblock {\em arXiv preprint arXiv:1711.05189}, 2017.

\bibitem{jang2022privacy}
Jaehee Jang, Younho Lee, Andrey Kim, Byunggook Na, Donggeon Yhee, Byounghan Lee, Jung~Hee Cheon, and Sungroh Yoon.
\newblock Privacy-preserving deep sequential model with matrix homomorphic encryption.
\newblock In {\em Proceedings of the 2022 ACM on Asia Conference on Computer and Communications Security}, pages 377--391, 2022.

\bibitem{kaissis2021end}
Georgios Kaissis, Alexander Ziller, Jonathan Passerat-Palmbach, Th{\'e}o Ryffel, Dmitrii Usynin, Andrew Trask, Ion{\'e}sio Lima~Jr, Jason Mancuso, Friederike Jungmann, Marc-Matthias Steinborn, et~al.
\newblock End-to-end privacy preserving deep learning on multi-institutional medical imaging.
\newblock {\em Nature Machine Intelligence}, 3(6):473--484, 2021.

\bibitem{kim2023optimized}
Dongwoo Kim and Cyril Guyot.
\newblock Optimized privacy-preserving cnn inference with fully homomorphic encryption.
\newblock {\em IEEE Transactions on Information Forensics and Security}, 18:2175--2187, 2023.

\bibitem{lee2021precise}
Junghyun Lee, Eunsang Lee, Joon-Woo Lee, Yongjune Kim, Young-Sik Kim, and Jong-Seon No.
\newblock Precise approximation of convolutional neural networks for homomorphically encrypted data.
\newblock {\em arXiv preprint arXiv:2105.10879}, 2021.

\bibitem{lou2021hemet}
Qian Lou and Lei Jiang.
\newblock Hemet: a homomorphic-encryption-friendly privacy-preserving mobile neural network architecture.
\newblock In {\em International conference on machine learning}, pages 7102--7110. PMLR, 2021.

\bibitem{lou2021safenet}
Qian Lou, Yilin Shen, Hongxia Jin, and Lei Jiang.
\newblock Safenet: A secure, accurate and fast neural network inference.
\newblock In {\em International Conference on Learning Representations}, 2021.

\bibitem{mishra2020delphi}
Pratyush Mishra, Ryan Lehmkuhl, Akshayaram Srinivasan, Wenting Zheng, and Raluca~Ada Popa.
\newblock Delphi: a cryptographic inference system for neural networks.
\newblock In {\em Proceedings of the 2020 Workshop on Privacy-Preserving Machine Learning in Practice}, pages 27--30, 2020.

\bibitem{podschwadt2020classification}
Robert Podschwadt and Daniel Takabi.
\newblock Classification of encrypted word embeddings using recurrent neural networks.
\newblock In {\em PrivateNLP@ WSDM}, pages 27--31, 2020.

\bibitem{ruder2016overview}
Sebastian Ruder.
\newblock An overview of gradient descent optimization algorithms.
\newblock {\em arXiv preprint arXiv:1609.04747}, 2016.

\bibitem{sealcrypto}
{M}icrosoft {SEAL} (release 3.5).
\newblock \url{https://github.com/Microsoft/SEAL}, April 2020.
\newblock Microsoft Research, Redmond, WA.

\bibitem{torralba200880}
Antonio Torralba, Rob Fergus, and William~T Freeman.
\newblock 80 million tiny images: A large data set for nonparametric object and scene recognition.
\newblock {\em IEEE transactions on pattern analysis and machine intelligence}, 30(11):1958--1970, 2008.

\bibitem{wang2018adversarial}
Yunhe Wang, Chang Xu, Chao Xu, and Dacheng Tao.
\newblock Adversarial learning of portable student networks.
\newblock In {\em Proceedings of the AAAI Conference on Artificial Intelligence}, volume~32, 2018.

\bibitem{wu2018ppolynets}
Wei Wu, Jian Liu, Huimei Wang, Fengyi Tang, and Ming Xian.
\newblock Ppolynets: Achieving high prediction accuracy and efficiency with parametric polynomial activations.
\newblock {\em IEEE Access}, 6:72814--72823, 2018.

\end{thebibliography}

\end{document}